# Spin-transfer-driven ferromagnetic resonance of individual nanomagnets

J. C. Sankey, P. M. Braganca, A. G. F. Garcia, I. N. Krivorotov, R. A. Buhrman, and D. C. Ralph

*Cornell University, Ithaca, NY, 14853 USA*

We demonstrate a technique that enables ferromagnetic resonance (FMR) measurements of the normal modes for magnetic excitations in individual nanoscale ferromagnets, smaller in volume by a factor of 1000 than can be probed by other methods. The measured peak shapes indicate two regimes of response: simple FMR and phase locking. Studies of the resonance frequencies, amplitudes, and linewidths as a function of microwave power, DC current, and magnetic field provide detailed new information about the exchange, damping, and spin-transfer torques that govern the dynamics in magnetic nanostructures.

Ferromagnetic resonance (FMR) is the primary technique for learning about the forces that determine the dynamical properties of magnetic materials. However, conventional FMR detection methods lack the sensitivity to measure individual sub-100-nm-scale devices that are of interest for a broad range of memory and signal-processing applications. For this reason, many new techniques are being pursued for probing magnetic dynamics on small scales, including Brillouin scattering [1] and FMR detected by Kerr microscopy [2], magnetic resonance force microscopy [3], X ray microscopy [4], and electrical techniques [5]. Nevertheless, the smallest isolated structure in which FMR (as distinct from electron spin resonance [6]) has been measured is 0.8 μm × 4.8 μm × 5 nm [5]. Here we demonstrate a simple new form of FMR that uses innovative methods both to drive and detect magnetic precession and thereby provides a detailed new understanding of the magnetic modes in individual nanomagnets. We excite precession not by applying an AC magnetic field as is done in other forms of FMR, but by using the spin-transfer torque from a spin-polarized AC current [7,8]. We detect the resulting magnetic motions electrically by using the precessing magnet as a mixer to rectify the applied microwave signal. We demonstrate detailed studies of FMR in single nanomagnets as small as $30 \times 90 \times 5.5 \, \text{nm}^3$. The method should be scalable to investigate fundamental physics in much smaller samples, as well. Our technique is similar to methods developed independently by Tulapurkar et al. [9] for radio-frequency detection, but we will demonstrate that the peak shapes measured there were not simple FMR.

We have achieved the following new results: (i) We measure magnetic normal modes of a single nanomagnet, including both the lowest-frequency fundamental mode and higher-order spatially non-uniform modes. (ii) By comparing the FMR spectrum to signals excited by a DC spin-polarized current, we demonstrate that different DC biases can drive different normal modes. (iii) From the resonance line shapes, we distinguish simple FMR from a regime of phase locking. (iv) From the resonance linewidths, we achieve efficient measurements of magnetic damping in a single nanomagnet.

Our samples have a nanopillar structure (Fig. 1(a), inset), consisting of two magnetic layers -- 20 nm of permalloy (Py = $Ni_{81}Fe_{19}$) and 5.5 nm of a $Py_{65}Cu_{35}$ alloy -- separated by a 12-nm copper spacer (see details in [10]). We pattern the layers to have approximately elliptical cross sections using ion milling. We focus here on one sample with cross section approximately $30 \times 90$ nm$^2$, but we also obtained similar results in $40 \times 120$ nm$^2$ and $100 \times 200$ nm$^2$ samples. We use different materials for the two magnetic layers so that by applying a perpendicular magnetic field $H$ we can induce an offset angle between their equilibrium moment directions (both the spin-transfer torque and the small-angle resistance response are zero otherwise). The room-temperature magnetoresistance (Fig. 1(a)) shows that the PyCu moment saturates out-of plane at $\mu_0 H \approx 0.3$ T, while the larger moment of Py does not saturate until approximately $\mu_0 H > 1$ T [11]. All of our FMR measurements are performed at low temperature ($\leq 10$ K), and the direction of $H$ is approximately perpendicular to the layers ($\hat{z}$ direction), tilted ~ 5º along the long axis of the ellipse ($\hat{x}$ direction) to control in-plane moment components. Positive currents correspond to electron flow from the PyCu to the Py layer. A diagram of our measurement circuit is shown in Fig. 1(b). Using a bias tee, we can apply current at both microwave frequencies ($I_{RF}\cos 2\pi ft$) and DC ($I_{DC}$) while measuring the DC voltage across the sample $V_{DC}$. If the frequency $f$ is set near a resonance of either magnetic layer, the layer will be driven to precess, producing a time-dependent resistance:

$$R(t) = R_0 + \Delta R(t) = R_0 + \text{Re}\left(\sum_{n=0}^{\infty} \Delta R_{nf} e^{in2\pi ft}\right), \quad (1)$$

where $\Delta R_{nf}$ can be complex. The voltage $V(t) = I(t)R(t)$ will contain a term involving mixing between $I_{RF}$ and $\Delta R(t)$, so that the measured DC voltage will be

$$V_{DC} = I_{DC}(R_0 + \Delta R_0) + \frac{1}{2} I_{RF} |\Delta R_f| \cos(\delta_f), \quad (2)$$

where $\delta_f$ is the phase of $\Delta R_f$. The final term enables measurement of spin-transfer-driven FMR. To reduce background signals and noise, we chop the microwave current bias at 1.5 kHz and measure the DC mixing signal $V_{\text{mix}} = V_{DC} - I_{DC}R_0$ using a lock-in amplifier.

In Fig. 1(c) we plot the FMR response $V_{\text{mix}}/I_{RF}^2$ measured for $I_{DC}$ near 0. We observe several resonances, appearing as either peaks or dips in $V_{\text{mix}}$. Small non-zero values of $I_{DC}$ can decrease the width of some resonances and make them easier to discern, as discussed below. By studying the resonances as a function of $H$, we can characterize the evolution of distinct normal modes (Fig. 1(d)). The largest resonances can be grouped in two sets, that we will call modes $A_0, A_1, A_2$ (filled symbols) and $B_0, B_1$ (open symbols), based on their $H$-dependence. Above $\mu_0 H = 0.3$ T, the field required to saturate the PyCu moment along $\hat{z}$, the frequencies of modes $A_0$, $A_1$, and $A_2$ shift in parallel, linearly with $H$, with the slope expected from the Kittel formula $df/dH = g\mu_B\mu_0/h$, with $g = 2.2 \pm 0.1$. As expected for the modes of a thin-film nanomagnet [12], the measured frequencies are shifted above the frequency for uniform precession of a bulk film, $f_{film} = (g\mu_B/h)[\mu_0 H - \mu_0 M_{\text{eff}}]$, with $\mu_0 M_{\text{eff}} = 0.3$ T. This $H$ dependence provides initial evidence that $A_0$, $A_1$, and $A_2$ are magnetic modes of the PyCu layer (additional evidence is presented later). The other two large resonances, $B_0$ and $B_1$, also shift together, with a weaker dependence on *H*. This is the behavior expected for modes of the Py layer, because the values of *H* shown in Fig. 1(d) are not large enough to saturate the Py layers out of plane. In addition to these modes, we observe small signals (not shown in Fig. 1(d)) at twice the frequencies of the main modes and near frequency sums (modes C).

Based on comparisons to simulations [12,13] and the fact that the lowest-frequency resonances $A_0$ and $B_0$ produce the largest resistance signals, we propose that these two resonances correspond to the lowest-frequency normal mode of the PyCu and Py layer, respectively. This is the mode that should have the most spatially-uniform precession amplitude (albeit not exactly uniform) [12,13]. The higher-frequency resonances $A_1$, $A_2$, and $B_1$ must correspond to higher-order nonuniform modes. The observed frequencies and frequency intervals are in the range predicted for normal modes of similarly-shaped nanoscale samples [12,13].

Next we compare the FMR resonances to spontaneous precessional signals that can be excited by a DC spin-polarized current $I_{DC}$ alone ($I_{RF}$=0) [14,15]. The power spectral density of resistance oscillations for DC-driven excitations at 420 mT is shown in Fig. 2(a), as measured with a spectrum analyzer [14]. We examine $I_{DC} > 0$, which gives the sign of torque to drive excitations in the PyCu layer only, not the Py layer. A single sharp peak appears in the DC-driven spectral density above a critical current $I_c = 0.3$ mA, and moves to higher frequency with increasing $I_{DC}$. The gradual increase in frequency can be identified with an increasing precession angle, which decreases the average demagnetizing field along $\hat{z}$ [16]. At larger values of $I_{DC}$, we observe additional peaks at higher *f* and switching of the precession frequency between different values, similar to the results of previous measurements [14-16] that have not been well explained before.

The FMR resonances are displayed in Fig. 2(b) at the same values of $I_{DC}$ shown in Fig. 2(a). We find that the FMR fundamental mode $A_0$ that we identified above with the PyCu layer is the mode that is excited at the threshold for DC-driven excitations. When $I_{DC}$ is large enough that the DC-driven mode begins to increase in frequency (585 µA), the shape of this FMR resonance changes from a simple Lorentzian peak to a more complicated structure with a dip at low frequency and a peak at high frequency. The FMR resonances $A_1$ and $A_2$ also vary strongly in peak shape and frequency as a function of positive $I_{DC}$, in a manner very similar to $A_0$, confirming that $A_1$ and $A_2$ (like $A_0$) are associated with the PyCu layer. The FMR modes $B_0$ and $B_1$ that we identified with the Py layer do not shift significantly in $f$ as a function of positive $I_{DC}$. This is expected, because positive $I_{DC}$ is the wrong sign to excite spin-transfer dynamics in the Py layer [7].

There has been significant debate about whether the magnetic modes which contribute to the DC-spin-transfer-driven precessional signals correspond to approximately uniform macrospin precession or to nonuniform spin-wave instabilities [17-20]. Our FMR measurements show directly that, at $I_c$, the DC-driven peak frequency is equal to that of the lowest-frequency RF-driven mode, the one expected to be most spatially uniform [12]. Higher values of $I_{DC}$ can also excite the spatially non-uniform mode $A_1$ and even produce mode-hopping so that mode $A_1$ can be excited when mode $A_0$ is not.

In order to analyze the FMR peak shapes, we make the simplifying assumption that the lowest-frequency modes $A_0$ and $B_0$ can be approximated by a macrospin model, with the Slonczewski form of the spin-transfer torque [7]. When the magnetic moments are initially at rest and $I_{RF}$ is applied to excite FMR, the resulting small-amplitude resonance is predicted [10] to have a simple Lorentzian lineshape

$$V_{mix}(f) \propto \frac{I_{RF}^2/\Delta_0}{1+[(f-f_0)/\Delta_0]^2}. \tag{3}$$

Here $f_0$ is the unforced precession frequency. The width $\Delta_0$ predicted for the PyCu layer in our experimental geometry is, to within 1% error for $\mu_0 H > 0.5$ T [10],

$$\Delta_0 = \alpha f_0, \tag{4}$$

where $\alpha$ is the Gilbert damping parameter. The measured FMR peak for mode $A_0$ at $I_{DC}=0$, for sufficiently small values of $I_{RF}$, is fit accurately by a Lorentzian, the amplitude scales $V_{mix} \propto I_{RF}^2$, and the width is independent of $I_{RF}$, as predicted by Eq. (3) (Figs. 3(a) and 4(a)). For $I_{RF} > 0.35$ mA, the peak eventually shifts to higher frequency and the shape becomes asymmetric, familiar properties for nonlinear oscillators [21]. From the magnitude of the frequency shift in similar signals (Fig. 3(b), inset), we estimate that the largest precession angle we have achieved is approximately 40°.

The peak shape for mode $B_0$ is also to good accuracy Lorentzian for small $I_{DC}$, but with negative sign. This sign is expected because when the Py moment precesses in resonance, positive current pushes the Py moment angle closer to the PyCu moment, giving a negative resistance response. The FMR peak shapes for the higher-order modes $A_1$, $A_2$, and $B_1$ are not as well-fit by Lorentzians. We plot the spectrum of DC-driven excitations for $I_{DC} = 0.52$ mA, $I_{RF} = 0$ in Fig. 3(b). The width is much narrower than the FMR spectrum for the same mode (inset), confirming arguments that the linewidths in these two types of measurements are determined by different physics [22].

We noted above that the FMR peak shape changes from a Lorentzian to a more complex shape for sufficiently large values of $I_{DC}$. (See the detailed resonance shapes in Fig. 3(b-c).) This shape change can be explained as a consequence of phase locking between $I_{RF}$ and the large-amplitude precession excited by $I_{DC}$ [23-26]. When the precession frequency increases with precession amplitude, the RF current can force the amplitude on the low-$f$ side of the resonance to be smaller than the equilibrium DC-driven trajectory. Under these conditions, the precession phase-locks approximately out of phase with the applied RF current ($\delta_f \approx 180°$), giving negative values of $V_{mix}$. RF frequencies on the high-$f$ side of the resonance produce phase-locking approximately in-phase with the drive and a positive $V_{mix}$. We have confirmed this picture by numerical integration of the macrospin model (Fig. 3(d)) [10]. Recently, Tulapurkar et al. [9] measured similar peak shapes, and proposed that they were caused by simple FMR with a torque mechanism different from the Slonczewski theory. We suggest instead that the peak shapes in [9] are due either to phase-locking to thermally-excited precession at room temperature (rather than simple FMR), or to the superposition of two FMR signals from different layers (one positive signal like that of $A_0$ and one negative like $B_0$).

A benefit of measuring the Lorentzian lineshape of simple FMR is that the linewidth allows a measurement of the magnetic damping $\alpha$, using Eq. (4). It is highly desirable to minimize the damping in spin-transfer-driven memory devices so as to decrease the current needed for switching [7]. Previously, $\alpha$ in magnetic nanostructures could only be estimated by indirect means [27,28]. As shown in Fig. 4(b), for $I_{DC} = 0$ we measure $\alpha = 0.040 \pm 0.001$ for the PyCu layer. This is larger than the damping for $Py_{65}Cu_{35}$ films in identically-prepared large-area multilayers as measured by conventional FMR, $\alpha_{film} = 0.021 \pm 0.003$. The cause of the extra damping in our nanopillars is not known, but it may be related to oxidation along the sides of the device [29]. As a function of increasing $I_{DC}$, the theory of spin-transfer torques predicts that the effective damping should decrease linearly, going to zero at the threshold for the excitation of DC-driven precession [7]. This is precisely what we find for mode $A_0$ (Fig. 4(b)). In contrast, the linewidth of mode $B_0$ decreases with decreasing $I_{DC}$. This is as expected for a Py-layer mode, because the sign of the spin-transfer torque should promote DC-driven precession in the Py layer at negative, not positive, $I_{DC}$.

We have demonstrated that spin-transfer-driven FMR measurements provide detailed information about the dynamics of magnetic normal modes in single 100-nm-scale magnetic samples. This technique will be of immediate utility in understanding and optimizing magnetic dynamics in nanostructures used for memory and microwave signal processing applications. Furthermore, both spin-transfer torques and magnetoresistance measurements become increasingly effective on smaller size scales. The same technique may therefore enable new fundamental studies of even smaller magnetic samples, approaching the molecular limit.

We thank J-M. L. Beaujour, A. D. Kent, and R. D. McMichael for performing FMR measurements on our bulk multilayers. We acknowledge support from the Army Research Office and from the NSF/NSEC program through the Cornell Center for Nanoscale Systems. We also acknowledge NSF support through use of the Cornell Nanofabrication Facility/NNIN and the Cornell Center for Materials Research facilities.

# Figure 1

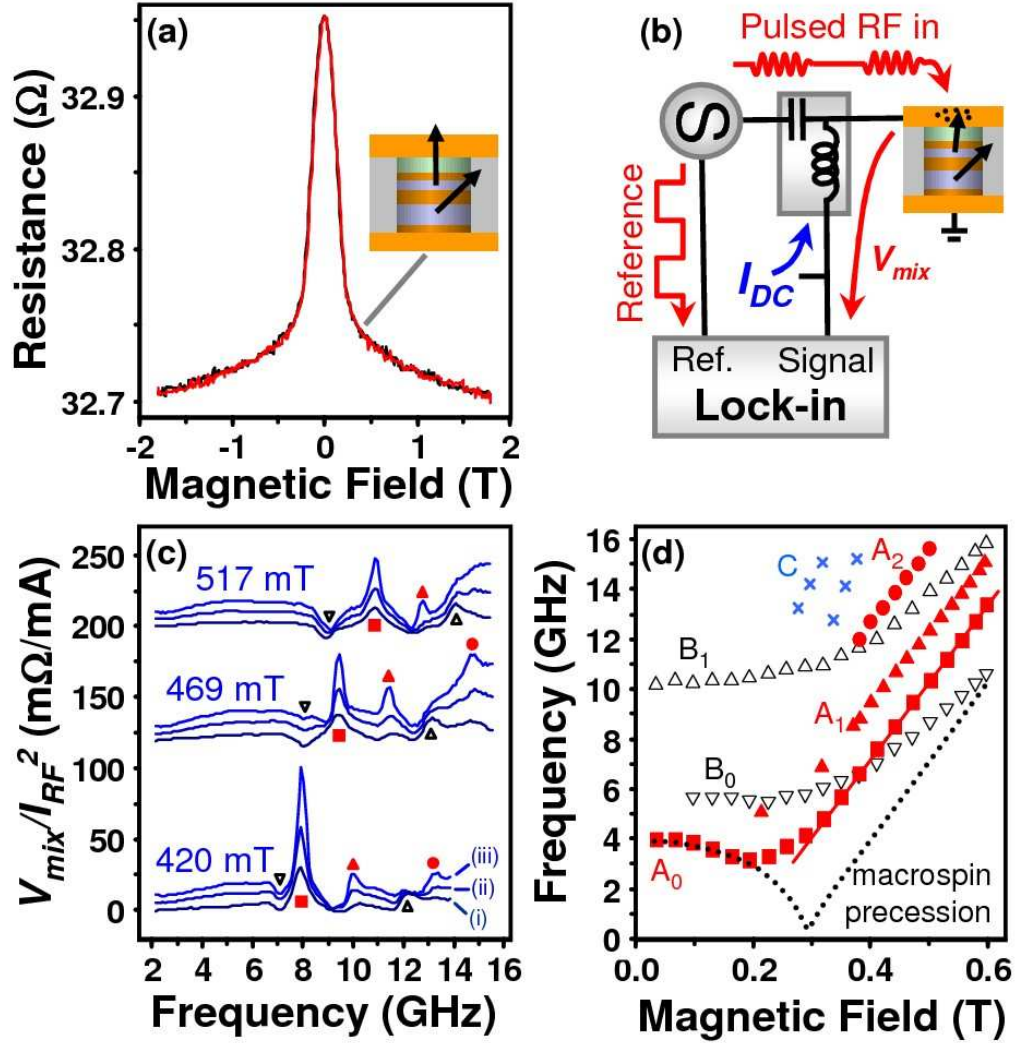

FIG. 1.(a) Room-temperature magnetoresistance as a function of field perpendicular to the sample plane. (inset) Cross-sectional sample schematic, with arrows denoting a typical equilibrium moment configuration in a perpendicular field. (b) Schematic of circuit used for FMR measurements. (c) FMR spectra measured at several values of magnetic field, at $I_{DC}$ values (i) 0, (ii) 150 μA, and (iii) 300 μA, offset vertically. Symbols identify the magnetic modes plotted in (d). Here $I_{RF}$ = 300 μA at 5 GHz and decreases by ~50% as $f$ increases to 15 GHz [10] (d) Field dependence of the modes in the FMR spectra. The solid line is a linear fit, and the dotted line would be the frequency of completely uniform precession.

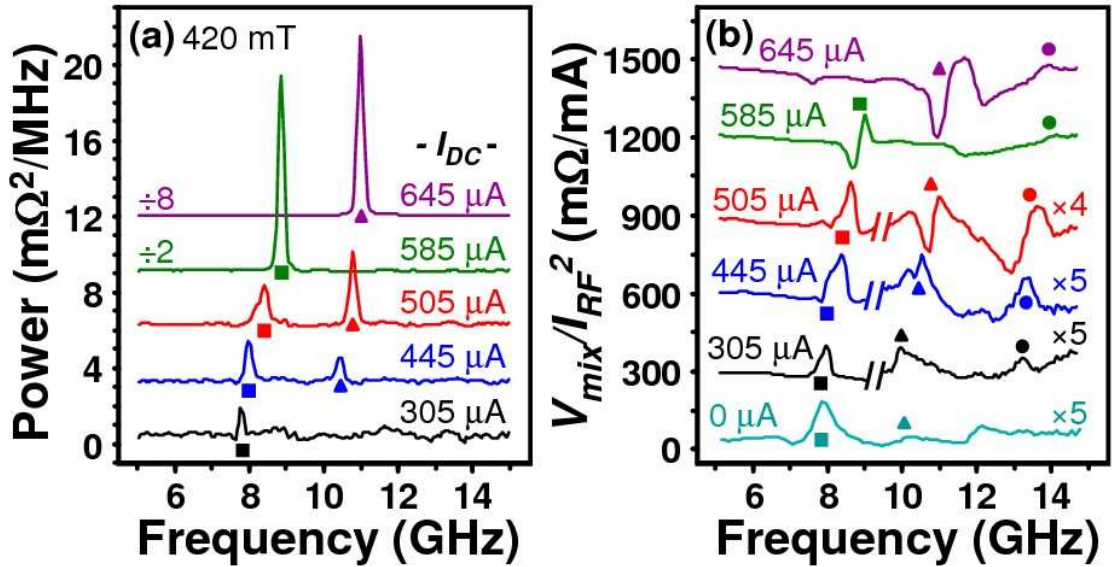

FIG. 2. Comparison of FMR spectra to DC-driven precessional modes. (a) Spectral density of DC-driven resistance oscillations for different values of $I_{DC}$ (labeled), with $\mu_0 H$ = 370 mT and $I_{RF}$ = 0. (b) FMR spectra at the same values of $I_{DC}$, measured with $I_{RF}$ = 270 μA at 10 GHz. The high-$f$ portions of the 305 μA, 445 μA, and 505 μA traces are amplified to better show small resonances. The $I_{DC}$=0 curve is the same as in Fig. 1(c).

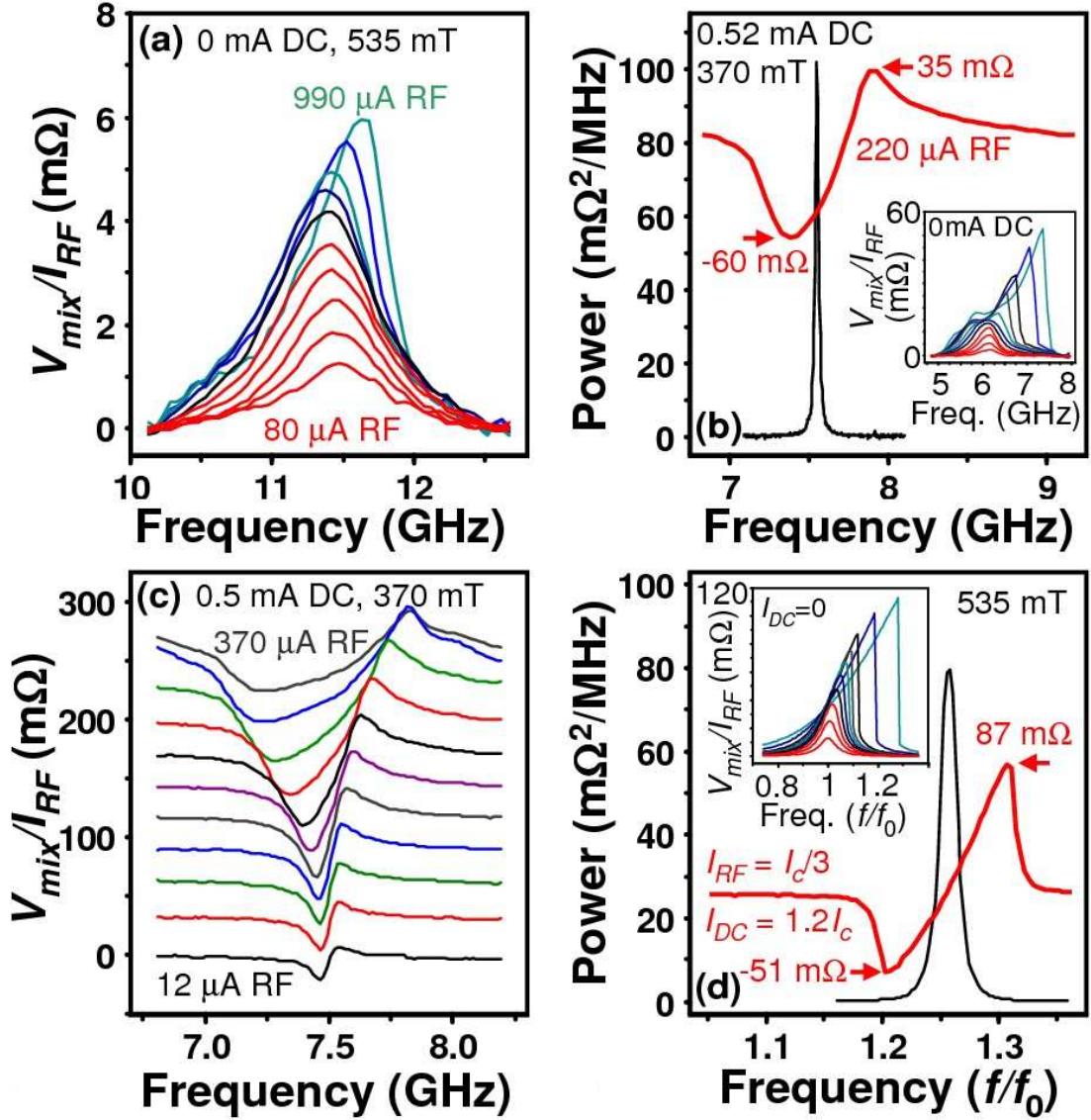

FIG. 3. (a) FMR peak shape for mode $A_0$ at $I_{DC} = 0$ and different values of $I_{RF}$: from bottom to top, traces 1-5 span $I_{RF} = 80 – 340$ μA in equal increments, and traces 5-10 span $340 – 990$ μA in equal increments. (b) Bottom curve: spectral density of DC-driven resistance oscillations for mode $A_0$, showing a peak with a half-width at half maximum = 13 MHz. Top curve: FMR signal at the same bias conditions, showing the phase-locking peak shape. (inset) Evolution of the FMR peak for mode $A_0$ at 370 mT, $I_{DC} = 0$, for $I_{RF}$

from 30 μA to 1160 μA. (c) Evolution of the FMR signal for mode $A_0$ in the phase-locking regime at $I_{DC}$ = 0.5 mA $\mu_0 H$ = 370 mT, for (bottom to top) $I_{RF}$ from 12 to 370 μA, equally spaced on a logarithmic scale. (d) Results of macrospin simulations for the DC-driven dynamics and the FMR signal.

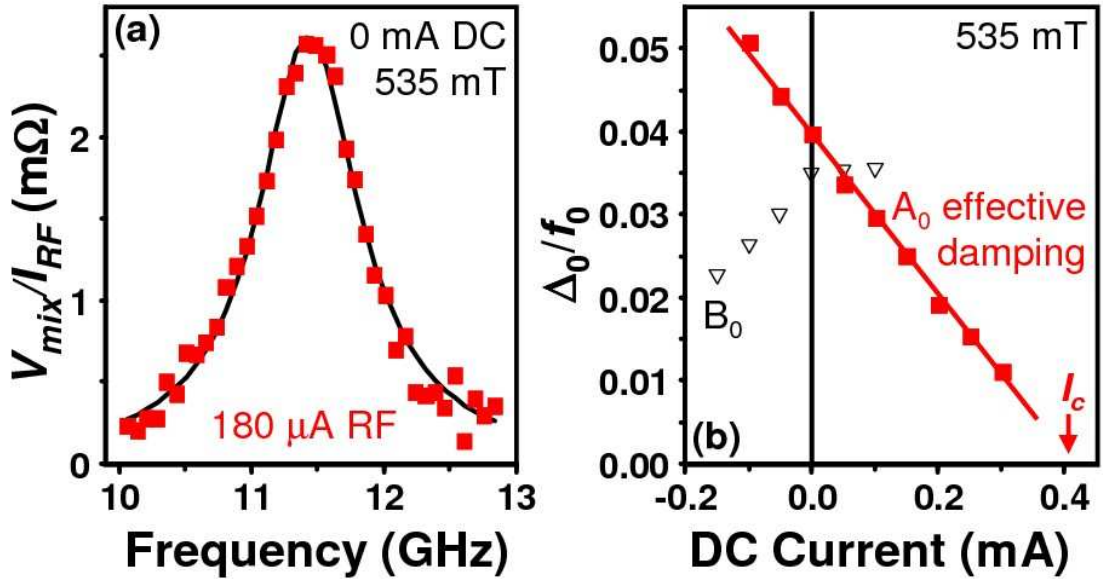

FIG. 4. (a) Detail of the peak shape for mode $A_0$, at $I_{DC}$ = 0, $I_{RF}$ = 180 μA, $\mu_0 H$ = 535 mT, with a fit to a Lorentzian lineshape. (b) Dependence of linewidth/(resonance frequency) on $I_{DC}$ for modes $A_0$ and $B_0$, for $\mu_0 H$ = 535 mT. For the PyCu layer mode $A_0$, $\Delta_0 / f_0$ is equal to the magnetic damping $\alpha$. The critical current is $I_c$ = 0.40 ± 0.03 mA at $\mu_0 H$ = 535 mT, as measured independently by the onset of DC-driven resistance oscillations.

# Supporting Material

## I. Circuit Calibration and Data Analysis

The RF attenuation in our cables, the bias tee, and the ribbon bonds connecting to the sample is frequency dependent. In order to know the value of $I_{RF}$ at the sample, this attenuation must be calibrated. We calibrate the attenuation of the cables and bias tee by measuring their transmission with a network analyzer.

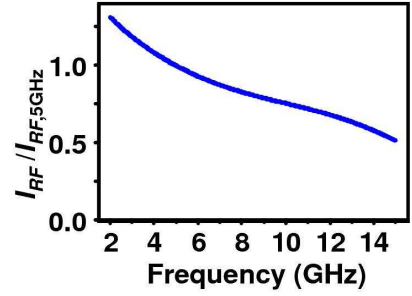

Figure S1

To estimate the losses due to the ribbon bonds, we measure the reflection from ribbon-bonded open, short, and 50-$\Omega$ test samples. We observe negligible reflection from the bonded 50 $\Omega$ sample, implying that the ribbon bonds produce little impedance discontinuity for frequencies < 15 GHz. We can therefore estimate the frequency-dependent transmission through the ribbon bonds as the square root of the measured reflection coefficient from either the bonded open test sample or the bonded short (a square root because the reflected power travels twice through the ribbon bonds). Finally, we measure the reflection coefficient directly for each of our ribbon-bonded samples before collecting FMR data, and from this determine its impedance and the resulting value of $I_{RF}$. For the $30 \times 90$ nm$^2$ sample on which we focus in the paper, the frequency dependence of $I_{RF}$ at the sample, referenced to the value at 5 GHz, is shown in Fig. S1.

The mixing signal contains a background due to deviations from linearity in the *I-V* curve of the sample, which we subtract from the data presented in the figures.

The thicknesses of the layers composing our samples are, from bottom to top, 120 nm Cu / 20 nm Py / 12 nm Cu / 5.5 nm Ni$_{81}$Fe$_{19}$ / 2 nm Cu / 30 nm Au, with a Au top contact. The difference in resistance between parallel and antiparallel magnetic layers for our $30 \times 90$ nm$^2$ sample at 10 K is $\Delta R_{max} = 0.84$ $\Omega$.

**II. Peak Shape Analysis for Spin-Transfer-Driven FMR**

In order to analyze our FMR peak shapes, we make the simplifying assumption that the lowest-frequency modes $A_0$ and $B_0$ can be approximated by a macrospin model, using the Landau-Lifshitz-Gilbert (LLG) equation of motion with the Slonczewski form of the spin-transfer torque [1]:

$$\frac{d\hat{m}}{dt} = \gamma\mu_0(\vec{H}+\vec{H}_{anis})\times\hat{m} + \alpha\hat{m}\times\frac{d\hat{m}}{dt} + c\frac{\eta I(t)}{e}\hat{m}\times(\hat{m}\times\hat{M}). \tag{S1}$$

Here $\hat{m}$ describes the moment direction of the precessing magnetic layer, $\gamma$ is the gyromagnetic ratio, $\vec{H}_{anis}$ accounts for shape anisotropy, $\alpha$ is the Gilbert damping, $\eta$ is a dimensionless efficiency factor, $\hat{M}$ is the moment direction of the static layer, and $c = +1$ for precession of the PyCu layer and $-1$ for precession of the Py layer. We consider the case of small-angle precession of the PyCu moment about $\hat{z}$. When $\hat{m}$ is initially at rest and $I_{RF}$ is applied to excite FMR, Eq. (3) predicts that the resulting resonance is Lorentzian

$$V_{mix}(f) = \frac{\eta c I_{RF}^2 \Delta R_{max} \sin^2(\theta_{stat})}{16\pi\Delta_0 e}\left(\frac{1}{1+[(f-f_0)/\Delta_0]^2}\right), \tag{S2}$$

where $\theta_{stat}$ is the angle between $\hat{M}$ and the precession axis, $f_0$ is the unforced precession frequency, and the width $\Delta_0$ is

$$\frac{\Delta_0}{f_0} = \alpha \frac{H/M_s - N_z + N_x/2 + N_y/2}{\sqrt{(H/M_s - N_z + N_x)(H/M_s - N_z + N_y)}}. \tag{S3}$$

We estimate that the effective demagnetization factors for our PyCu layer are $N_z = 0.797$, $N_x = 0.027$, and $N_y = 0.176$, based on a magnetization of 0.39 T [2] and coercive field measurements. However, the result of Eq. (5) is quite insensitive to these values, so that for $\mu_0 H > 0.5$ T we have simply $\Delta_0/f_0 = \alpha$ for the PyCu layer to within 1% error. Simulations show that this prediction is also not altered at the 1% level by the 5° offset between $\vec{H}$ and the $\hat{z}$ direction in our measurements.

For the Py layer mode, there is an additional correction required to relate $\Delta_0/f_0$ to $\alpha$, due to the larger deviation of the precession axis from $\hat{z}$.

### III. Simulation Parameters

In our numerical simulations, we integrate the LLG equation for macrospin precession (Eq. (S1)), using the following parameters: $\alpha = 0.04$, $g = 2.2$, a PyCu magnetization $\mu_0 M_s = 390$ mT [2], in- and out-of-plane anisotropies 58 mT and 300 mT, and an efficiency parameter $\eta = (0.2)g\mu_B/(2M_s V)$, where $\mu_B$ is the Bohr magneton and $V$ is the volume of a 5.5-nm-thick disk of elliptical cross section $90 \times 30$ nm$^2$. Thermal effects are modeled with a 10 K Langevin fluctuating field [3]. The qualitative results of the simulation are not affected by reasonable variations in device parameters.

**IV. Regarding another proposed mechanism for DC voltages produced by magnetic precession:**

Berger has proposed that a precessing magnet in a multilayer device may generate a DC voltage directly [4]. This mechanism would produce another source of signal in our experiments on resonance, in addition to the mixing mechanism we discussed in the main text. However, the maximum magnitude of $V_{DC}$ predicted to be generated by the Berger mechanism is $hf/e = 40$ $\mu$eV for $f = 10$ GHz, and our FMR signals can grow much larger than this. Also, we find that at small values of $I_{RF}$ our signals scale as $V_{DC} \propto I_{RF}^2$ as expected for the mixing mechanism (because $|\Delta R_f| \propto I_{RF}$), while the Berger signal would scale $\propto I_{RF}$. On this basis, we argue the mixing mechanism is dominant in producing our signal, and we have considered only this mechanism in our analysis.